\documentclass[3p,times,twocolumn]{elsarticle}

\usepackage{ecrc}

\volume{00}

\firstpage{1}

\journalname{Nuclear Physics B Proceedings Supplement}

\runauth{T. Hurth, F. Mahmoudi}

\jid{nuphbp}

\jnltitlelogo{Nuclear Physics B Proceedings Supplement}

\usepackage{amssymb}

\biboptions{sort&compress}

\usepackage[figuresright]{rotating}
\input{definitions.tex}

\begin{document}

\begin{frontmatter}




\title{Signs for new physics in the recent LHCb data?\tnoteref{label1}}  
\tnotetext[label1]{Based on talks given by T.H. at the Fifth Workshop on Theory, Phenomenology and Experiments in Flavour Physics, Capri,  23-25 May 2014 and
at the FPCP 2014 conference on Flavor Physics and CP Violation, Marseille, 25-30 May 2014.}


\author{Tobias Hurth}
\address{PRISMA Cluster of Excellence, Institute for Physics (THEP)\\
Johannes Gutenberg University, D-55099 Mainz, Germany}
\ead{tobias.hurth@cern.ch}

\author{Farvah Mahmoudi}
\address{Universit{\' e} de Lyon, Universit{\' e} Lyon 1, F-69622 Villeurbanne Cedex, France;\\
Centre de Recherche Astrophysique de Lyon, Saint Genis Laval Cedex, F-69561, France;\\ CNRS, UMR 5574;
Ecole Normale Sup{\' e}rieure de Lyon, France}
\address{CERN Theory Division, Physics Department, CH-1211 Geneva 23, Switzerland}
\ead{nazila@cern.ch}

\begin{abstract}
We comment on  some tensions with the Standard Model predictions in the recent LHCb data. 
\end{abstract}




\end{frontmatter}



\section{Introduction}

With the first measurement of new angular observables in the exclusive decay $B \to K^*\mu^+\mu^-$ based on the 1~fb$^{-1}$ dataset, LHCb has found  a 4.0$\sigma$ local discrepancy  in one of the $q^2$ bins for one of the angular observables~\cite{Aaij:2013qta}, namely in the bin $q^2 \in [ 4.3,8.63 ]$ GeV$^2$ of the observable $P_5^{'}$.
The latter belongs to the set of optimised observables in which form factor dependence cancels out to first order. LHCb results are compared here with the theoretical predictions in Ref.~\cite{Descotes-Genon:2013vna}.  Intriguingly, other smaller but consistent deviations are also present in other observables~\cite{Aaij:2013qta}.  

In the low-$q^2$ region, the up-to-date description of exclusive heavy-to-light $B \to K^* \mu^+\mu^-$ decays is the method of QCD-improved Factorisation (QCDF) and its field-theoretical formulation of Soft-Collinear Effective Theory (SCET). In the combined limit of a heavy $b$-quark and of an energetic $K^*$ meson,  the decay amplitude factorises to leading order in $\Lambda/m_b$ and to all orders in $\alpha_s$ into process-independent non-perturbative quantities like $B\to K^*$ form factors and light-cone distribution amplitudes (LCDAs) of the heavy (light) mesons and perturbatively calculable quantities, which are known to
$O(\alpha_s^1)$~\cite{Beneke:2001at,Beneke:2004dp}. 
Further, the seven a priori independent $B\to K^*$ QCD form factors reduce to two universal soft form factors $\xi_{\bot,\|}$~\cite{Charles:1998dr}. 
The factorisation formula applies well in the dilepton mass range $1\; {\rm GeV}^2 < q^2 < 6\; {\rm GeV}^2$.

Taking into account all these  simplifications the various \Kstar spin amplitudes at leading order in $\lqcd/m_b$ and \as turn out to be linear in the soft form factors $\xi_{\bot,\|}$ and also in the short-distance Wilson coefficients. As was explicitly shown in Refs.~\cite{Egede:2008uy, Egede:2010zc}, these simplifications allow to design a set of optimised observables, in which any soft form factor dependence (and its corresponding uncertainty) cancels out for all low dilepton mass squared~\qsq at leading order in \as and $\lqcd/m_b$. An optimised set of independent observables  was constructed in Refs.~\cite{Matias:2012xw,Descotes-Genon:2013vna}, in which almost all  observables  are free from hadronic uncertainties which are related to the form factors.

However, the soft form factors  are {\it not} the only source of hadronic uncertainties in these angular observables. 
It is well-known that within the QCDF/SCET approach, a general, quantitative method to estimate the important $\lqcd/m_b$ corrections to the heavy quark limit is missing. 
It is clear that the interpretation of the LHC measurement strongly depends on the treatment of this problem as we discuss in the next section.

There is another issue: The validity of the theory predictions based on QCD factorisation approach within the region $q^2 \in [4.3, 8.63]$ GeV$^2$ is highly questionable. 
The validity is commonly assumed up to 6 GeV$^2$  for two reasons. The perturbative description of the charm loops is valid in this region and also the kinematical assumptions about the large energy of the $K^*$ within the SCET/QCD factorisation approach are still reasonable. Thus, using the theory predictions up to 8.63 GeV$^2$ could induce larger hadronic corrections.

Leaving these two issues aside,  it has been shown that the deviation in the observable $P_5^{\prime}$ and the small deviations in other  observables in the low-$q^2$ area, can be  consistently described by a smaller $C_9$ Wilson coefficient, together with a less significant contribution of a non-zero $C_9^{\prime}$ (see for example Ref.~\cite{Descotes-Genon:2013zva}). 
This is a challenge for the model-building as we will discuss below. 

Thus, it is not clear if the anomaly is a sign for new physics beyond the SM, or a consequence of underestimated hadronic power corrections or non-perturbative charm effects or just  a statistical fluctuation. The LHCb analysis based on the 3 fb$^{-1}$ dataset is eagerly awaited to clarify the situation.

More recently, another small discrepancy occurred. The ratio $R_K = {\rm BR}(B^+ \to K^+ \mu^+ \mu^-) / {\rm BR}(B^+ \to K^+ e^+ e^-)$ in the low-$q^2$ region has been
measured by LHCb showing a $2.6\sigma$ deviation from the SM prediction~\cite{Aaij:2014ora}. 
In contrast to the anomaly in the rare decay $B \rightarrow K^{*} \mu^+\mu^-$ which is affected by unknown power corrections, the ratio $R_K$ is theoretically rather clean. This might be a sign for lepton non-universality.

\section{Power corrections} 

In spite of the fact that the power corrections cannot be calculated, the corresponding uncertainties should be made manifest within the theory predictions. Therefore, in Refs.~\cite{Egede:2008uy,Egede:2010zc} the effects of the $\Lambda_{\rm QCD}/m_{b}$ corrections has been parametrised for each of the \Kstarz spin-amplitudes with some {\it unknown} linear correction. In case of CP-conserving observables this just means  $A'_{i} =  A_{i}(1 + C_{i})$, where $C_{i}$ is the relative amplitude. In the case of CP-violating observables, a strong phase has to be included (see Ref.~\cite{Egede:2010zc} for details).  It is further assumed that these amplitudes ($C_{i}$) are not functions of $q^2$, although in practice they may actually be, and any unknown correlations are  also ignored. 

Based on a simple dimensional estimate, one has chosen $|C_i| < 10\%$. 
There are also {soft}  arguments for this choice:
Under the assumption that the main part of the $\lqcd/m_b$ corrections is included in the full form factors, the difference of the theoretical results using the full QCD form factors on one hand and the soft form factors on the other hand confirms this simple dimensional estimate. In fact, the comparison of the approaches leads to a $7\%$ shift of the central value {\it at the level of observables}. 
Secondly, one can state that the chiral enhancement of $\Lambda_{\rm QCD}/m_\b$ corrections in the case of hadronic $B$ decays does not occur in the case of the semileptonic decay mode with a {\it vector} final state.
Thus, it is not expected that they are as large as $20-30\%$ as in the $B \to \pi\pi$ decay.

This procedure was introduced to make the  unknown $\Lambda_{\rm QCD}/m_{b}$ corrections manifest; this ansatz, put in by hand, was never meant as a real quantitative estimate. In fact assuming $10\%$ corrections on the amplitude level one often finds  power corrections of less than $5\%$ on the observable level as one can read off from Table 5 and 6 of Ref.~\cite{Descotes-Genon:2013vna}, in which the hand-made error due to $\Lambda_{\rm QCD}/m_{b}$ corrections is nicely separated from the error due to input parameters and scale dependence. 
So the errors due to the power corrections given in Ref.~\cite{Descotes-Genon:2013vna} are most probably an underestimation of the hadronic uncertainty if taken as  a real quantitative estimate. However, if we assume $10\%$ error due to the unknown power corrections -- which corresponds to a naive dimension estimate of $\Lambda/m_b$ on the level of observables and is also backed up by some soft arguments (see above) -- we find the pull in case of the third bin of the observable $P_5^{'}$ reduced from $4.0\sigma$ to $3.6\sigma$ what still represents a significant deviation. And even if one assumes $30\%$ error then the pull in this case is still $2.2\sigma$ within the model-independent analysis presented in Ref.~\cite{Descotes-Genon:2013wba}.
\begin{figure*}[!t]
\begin{center}
\includegraphics[width=5.5cm]{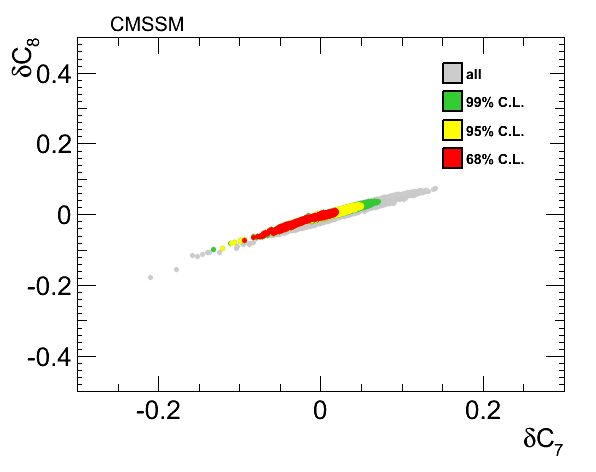}\includegraphics[width=5.5cm]{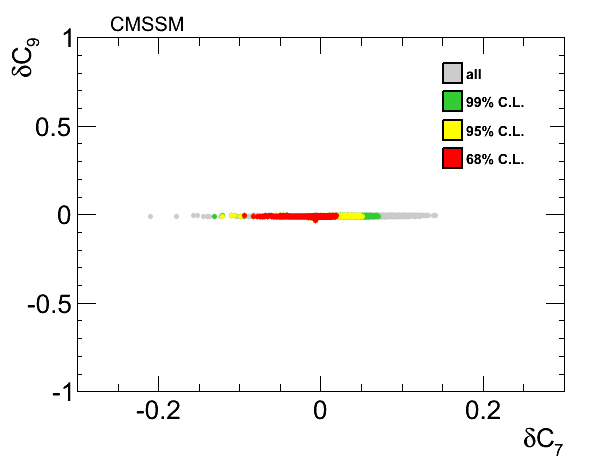}\includegraphics[width=5.5cm]{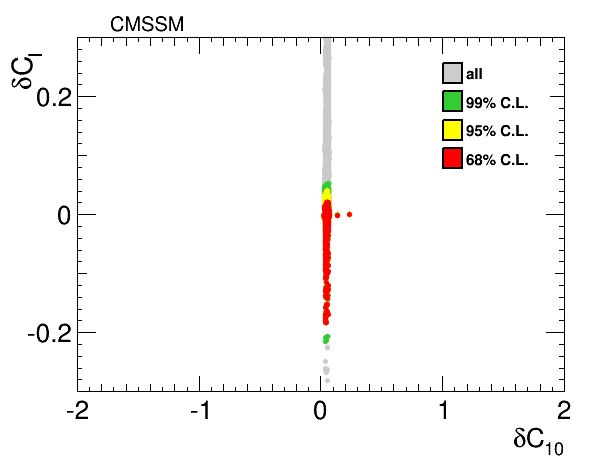}\\
\includegraphics[width=5.5cm]{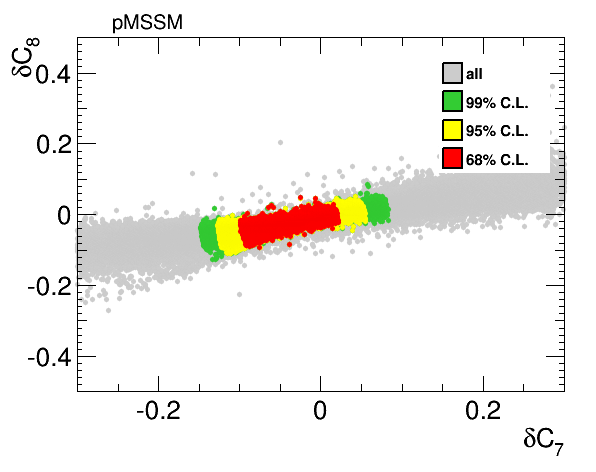}\includegraphics[width=5.5cm]{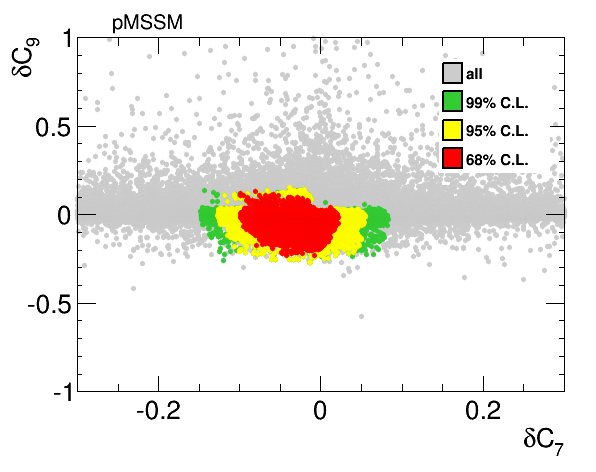}\includegraphics[width=5.5cm]{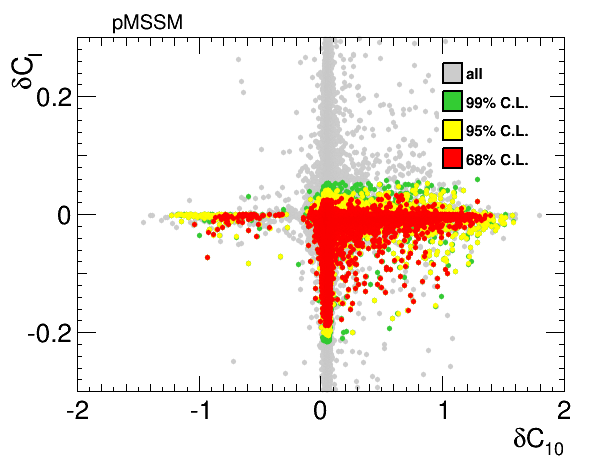}
\caption{Global fit to the NP coefficients $\delta \C{i}$ at the $\mu_b$ scale in CMSSM (upper row) and in pMSSM (lower row), at 1$\sigma$ (red), 2$\sigma$ (yellow) and 3$\sigma$ (green) imposing all the flavour observables in addition to the Higgs mass constraint.}
\label{fig:fit-mH}
\end{center}
\end{figure*}

In Ref.~\cite{Jager:2012uw} a general parametrisation for the power corrections to the form factor terms (the factorisable piece in QCD factorisation) is given,  
There are two free parameters in the ansatz for each QCD form factor which have to be determined. 
In a more recent analysis~\cite{Descotes-Genon:2014uoa}, this procedure gets further developed. 
The authors analyse the  factorisable power \mbox{corrections} by looking at the difference of QCD form factors and soft form factors, an argument that was already used to get an order of magnitude estimate of power corrections. Unfortunately, the errors of the light-cone sum rule calculations of QCD form factors are too large and the correlations between the various sum rule calculations are not commonly known in order to get reasonable results. Thus, in Ref.~\cite{Descotes-Genon:2014uoa}  only the central values get fixed by this procedure and a $10\%$ error is attributed to the power corrections by hand again.  
Another argument is put forward that one can choose the renormalisation scheme of the soft form factors such that the power corrections to certain observables are reduced. 
In addition one has to estimate the non-factorisable power corrections. Here an observation is helpful~\cite{Altmannshofer:2008dz}, that non-factorisable corrections are not induced by the leading  electromagnetic and semileptonic operators. However, also these contributions have to be estimated by a hand-made ansatz; this is done in Ref.~\cite{Descotes-Genon:2014uoa} by a $q^2$ dependent correction of order $10\%$ on the amplitude level. 
Clearly, an improved, but honest light-cone sum rule calculation of the various QCD form factors including their correlations is really needed to make progress on this issue. But the non-factorisable power contributions will then be still a source of uncertainty limiting the new physics sensitivity of these exclusive decay modes. 

In principle, the light-cone sum rule (LCSR) approach allows to estimate these non-factorisable contributions (with the well-known uncertainties of QCD sum rules) as was demonstrated in the case of the decay $B \to K \ell^+\ell^-$~\cite{Khodjamirian:2012rm}.
For the $B \to K^*\ell^+\ell^-$ case, only soft-gluon contributions of the charm loop effects has been considered yet~\cite{Khodjamirian:2010vf}.

\section{New physics interpretations of the anomaly}

Consistent SM and new physics interpretations of the measured  deviation in the $B \to K^* \ell^+\ell^-$ mode have been discussed in a large number of 
references~\cite{Descotes-Genon:2013wba,Altmannshofer:2013foa,Hambrock:2013zya,Gauld:2013qba,Buras:2013qja,Gauld:2013qja,Datta:2013kja,Beaujean:2013soa,Buras:2013dea,Hurth:2013ssa,Mahmoudi:2014mja,Altmannshofer:2014cfa}. 
In a model-independent analysis, the anomaly can be consistently described by smaller $C_9$ and $C'_9$ Wilson coefficients. 
The usual suspects like the MSSM, warped extra dimension scenarios, or models with partial compositeness, cannot accommodate the deviation at the $1\sigma$ level, but $\mbox{Z}^{'}$models may do this~\cite{Gauld:2013qba}.  

In the MSSM, we have no means of generating any sizeable contribution to the coefficient $C'_9$, but also any significant contribution to $C_9$ is correlated to contributions to other Wilson coefficients affecting the other observables. Nevertheless, combining all the observables in a fit one can check the global agreement of the model with the available data~\cite{Mahmoudi:2014mja}. This is shown in Fig.~\ref{fig:fit-mH} for the relevant Wilson coefficients in the constrained MSSM (CMSSM) and in the more general setup of the phenomenological MSSM (pMSSM), where all the generated points are shown (grey), indicating the values for the Wilson coefficients reachable within the MSSM, as well as the points satisfying the global 1,2,3$\,\sigma$ constraints (red, yellow, green). The Higgs mass constraint has also been imposed. As can be seen, the overall agreement is fairly good, with regions in SUSY parameter space where the absolute $\chi^2$ is sufficiently small and an agreement at the 1$\sigma$ level is reached.

\section{Signs for lepton non-universality}

\begin{figure}[t!]
\begin{center}
\includegraphics[width=7.cm]{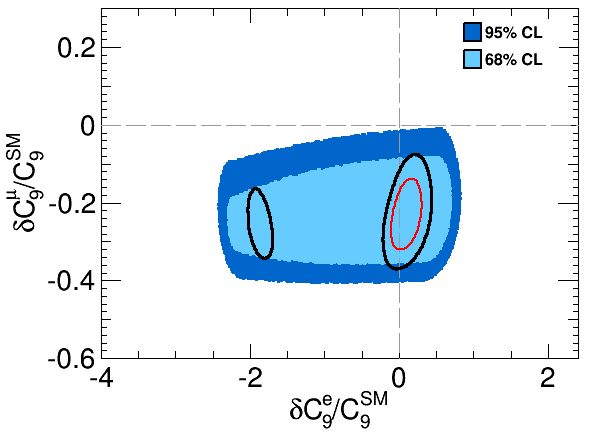}
\caption{Global fit results for $C_9^\mu, C_9^e, C_{10}^\mu, C_{10}^e$. The red and black contours correspond to the 1 and 2$\sigma$ regions respectively of the two operator only fit for $(C^e_9,C^\mu_9)$.\label{fig:c9c10}}
\end{center}
\end{figure}

\begin{figure}[t!]
\begin{center}
\includegraphics[width=7.cm]{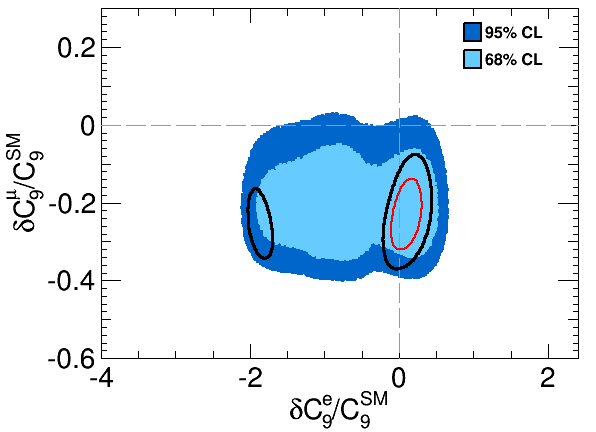}
\caption{Global fit results for $C_9^\mu, C_9^{'\mu}, C_9^e, C_9^{'e}$. The red and black  contours correspond to the 1 and 2$\sigma$ regions respectively of the two operator only fit for $(C^e_9,C^\mu_9)$.\label{fig:c9c9p}}
\end{center}
\end{figure}

Besides this known anomaly in the angular analysis of $B \rightarrow K^{*} \mu^+\mu^-$ decay, another small discrepancy recently occurred. The ratio $R_K = {\rm BR}(B^+ \to K^+ \mu^+ \mu^-) / {\rm BR}(B^+ \to K^+ e^+ e^-)$ in the low-$q^2$ region has been measured by LHCb showing a $2.6\sigma$ deviation from the SM prediction~\cite{Aaij:2014ora}. 
This discrepancy has been addressed in a few recent studies~\cite{Alonso:2014csa,Hiller:2014yaa,Ghosh:2014awa,Biswas:2014gga,Davidtalk:2014,Hurth:2014vma,Glashow:2014iga}.
A global fit to all observables considering separately new physics contributions to the electron and muon semileptonic Wilson coefficients $C_{9,10}^e$ and $C_{9,10}^\mu$ (and the corresponding chirality flipped coefficients) is shown in Figs.~\ref{fig:c9c10} and \ref{fig:c9c9p} (see \cite{Hurth:2014vma} for more detail). Fig.~\ref{fig:c9c10} shows the fit results for $C_9^\mu, C_9^e, C_{10}^\mu, C_{10}^e$, while Fig.~\ref{fig:c9c9p} presents the results for $C_9^\mu, C_9^{'\mu}, C_9^e, C_9^{'e}$. 
We see that the SM is disfavoured at the 2$\sigma$ level. Yet there is tension in the muon sector for $C_9$. 
In order to compare the results with the scan for two operators $\{O_9^\mu, O_9^e\}$ only, the contours corresponding to the 1 and 2$\sigma$ fit result for $\{C_9^\mu, C_9^e\}$ are overlaid in the figures. This shows that considering arbitrarily only two operators can be too restrictive and even misleading since a large area of new physics parameter space might be unjustifiably overlooked. For example, in the two-operator fit lepton-universality, $\delta C_9^\mu = \delta C_9^e$, is disfavoured by $2 \sigma$, while within the four-operator fit the agreement is improved.

\section{Cross-checks with the inclusive mode}

The inclusive mode  $B\to X_s \ell^+ \ell^-$ can only be measured at $e^+ e^-$ machines and is theoretically cleaner than the exclusive modes~\cite{Hurth:2010tk,Hurth:2012vp}. 
The theoretical accuracy in the low-$q^2$ region is of the order of $10\%$~\cite{Huber:2007vv}. But the branching fraction has been measured by Belle and BaBar using the sum-of-exclusive technique only.  

In Refs.~\cite{Hurth:2013ssa, Hurth:2014vma} we checked the compatibility of the present datasets of the exclusive and inclusive modes. It is remarkable result that the  sets of exclusion plots are nicely compatible with each other.  This is a non-trivial consistency check. 
At the moment, the measurements of the $B \rightarrow K^* \ell^+\ell^-$ observables are the most powerful ones. 
However, the latest published measurement of Belle~\cite{Iwasaki:2005sy}  is based on a sample of $152 \times 10^6$ $B \bar B$ events only, which corresponds to less than $30\%$ of the dataset available at the end of the Belle experiment while BaBar has just recently presented an analysis based on the whole dataset using a sample of $471 \times 10^6$ $B \bar B$ events~\cite{Lees:2013nxa} overwriting  the previous measurement from 2004 based on $89 \times 10^6$ $B \bar B$  events~\cite{Aubert:2004it}.
Moreover,  there will be a Super-$B$ factory Belle-II with a final integrated luminosity of 50 ab$^{-1}$~\cite{belle2}.
There is a recent analysis~\cite{Kevin2} of the expected total uncertainty on the partial decay width and the forward-backward asymmetry in several bins of dilepton mass-squared for the fully inclusive $B \to  X_s \ell^+\ell^-$ decays assuming a 50 ab$^{-1}$ total integrated luminosity (for details see Ref.~\cite{Hurth:2013ssa}).
One finds a relative fractional uncertainty of $2.9\%$  ($4.1\%$) for the branching fraction in the low- (high-)$q^2$ region and a total absolute uncertainty of 0.050 in the low-$q^2$ bin 1 ($1<q^2<3.5$ GeV$^2$), 0.054 in the low-$q^2$ bin 2 ($3.5<q^2<6$ GeV$^2$) and 0.058 in the high-$q^2$ interval ($q^2>14.4$ GeV$^2$)  for the {\it normalised}  $A_{FB}$.
So the inclusive mode will lead to very strong constraints on the Wilson coefficients

\begin{figure}[t!]
\begin{center}
\includegraphics[width=7.cm]{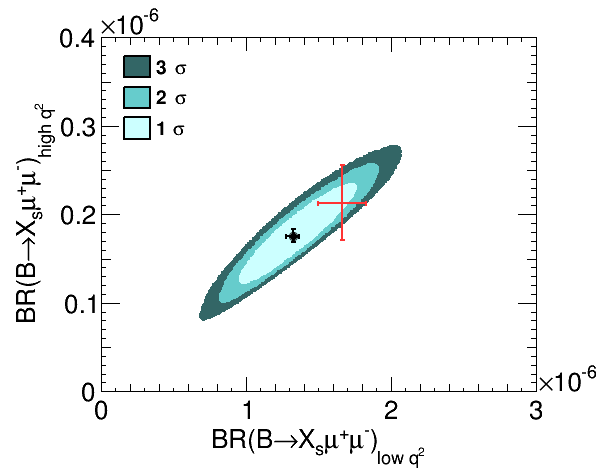}
\caption{1, 2 and 3$\sigma$ ranges for the branching ratio at low- and high-$q^2$ within the model-independent analysis. 
Future measurement at the high-luminosity Belle-II Super-$B$-Factory assuming the best-fit point of the model-independent analysis as central value (black) and the SM predictions (red/grey).\label{fig:bsllc}}
\end{center}
\end{figure}

\begin{figure}[t!]
\begin{center}
\includegraphics[width=7.cm]{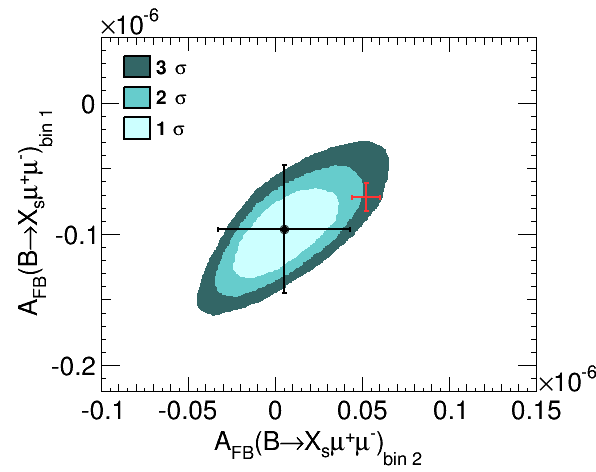}
\caption{1, 2 and 3$\sigma$ ranges for the {\it unnormalised} forward-backward asymmetry in bin 1 ($1<q^2<3.5$ GeV$^2$) and in bin 2 ($3.5<q^2<6$ GeV$^2$) within the model-independent analysis. Future measurement at the high-luminosity Belle-II Super-$B$-Factory assuming the best-fit point of the 
model-independent analysis as central value (black) and the SM predictions (red/grey).\label{fig:bslld}}
\end{center}
\end{figure}

We illustrate the usefulness of these  future measurements of the inclusive mode at Belle-II in the following way~\cite{Hurth:2014vma}.
We make a model independent fit for the coefficients $C_7$, $C_8$, $C_9$, $C_{10}$ and $C_{l}$ (for notation see Ref.~\cite{Hurth:2013ssa}). In addition to all the $b \to s \ell^+\ell^-$  observables, we consider the inclusive branching ratio of $B\to X_s \gamma$ as well as the isospin asymmetry in $B\to K^* \gamma$ \mbox{decay} which are relevant to constrain $C_7$ and $C_8$.  
Based on our model-independent analysis we predict the branching ratio at low- and high-$q^2$. In Fig.~\ref{fig:bsllc}, we show the 1, 2, and 3$\sigma$ ranges for these observables. In addition, we add the future measurements at Belle-II assuming the best fit solution of our model-independent analysis as central value. These measurements are indicated by the black error bars. They should be compared with the theoretical SM predictions given by the red error bars. 
Fig.~\ref{fig:bsllc} indicates that the future measurement of the inclusive branching ratios separates nicely from the SM prediction as the model-independent fit. And also the future measurement of the forward-backward asymmetry at Belle-II will allow us to separate the potential new physics measurement from the SM prediction in a significant way as shown in Fig.~\ref{fig:bslld}. 

%

\section{Global fit in MFV}

Assuming the tensions seen in the recent LHCb results are hints for new physics, it is important to consider whether new flavour structures are needed to explain the data.  
The hypothesis of MFV~\cite{Chivukula:1987py,Hall:1990ac,D'Ambrosio:2002ex,Hurth:2008jc,Hurth:2012jn} implies that flavour and CP symmetries are broken as in the SM. Thus, it requires that all flavour- and CP-violating interactions be  linked to the known structure of Yukawa couplings.     
The MFV hypothesis represents an important benchmark in the sense that any measurement which is inconsistent with the general constraints and relations induced by the MFV hypothesis unambiguously indicates the existence of new flavour structures. 

\begin{figure*}[!t]
\begin{center}
\includegraphics[width=5.5cm]{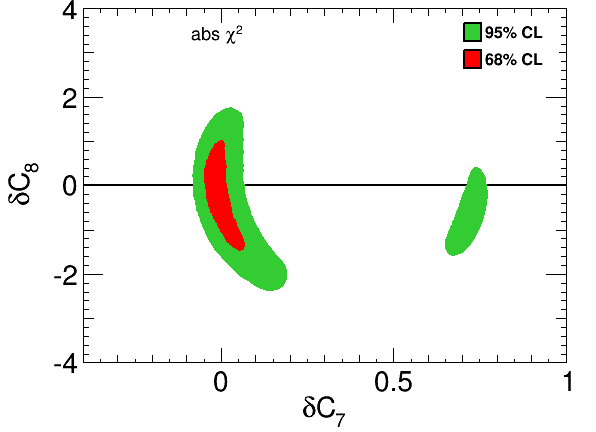}\includegraphics[width=5.5cm]{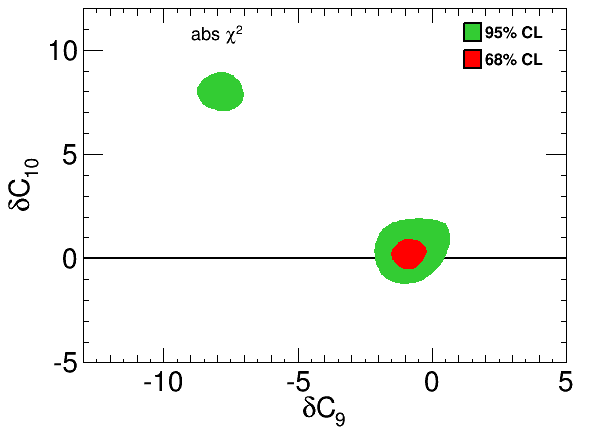}\includegraphics[width=5.5cm]{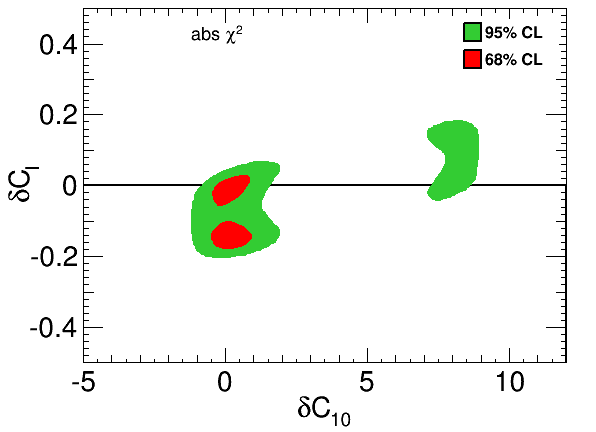}
\caption{Global fit to the NP contributions $\delta C_i$ in the MFV effective theory, at 1 (red) and 2$\sigma$ (green).}
\label{fig:fit-MFV}
\end{center}
\end{figure*}

We study the results of the global fit for the new physics contributions to the Wilson coefficients $C_7$, $C_8$, $C_9$, $C_{10}$ and $C_{l}$, and update the results of Refs. \cite{Hurth:2012jn,Hurth:2013ssa} based on the latest experimental results. As can be seen in Fig.~\ref{fig:fit-MFV}, while the 1 and 2$\sigma$ allowed regions are squeezed compared to those of \cite{Hurth:2012jn,Hurth:2013ssa} which shows the impact of the new measurements, the overall agreement of the MFV solutions with the data is still very good, and no new flavour structure is needed to explain the experimental results.

\section*{Acknowledgements}  

FM and TH thank the organisers for their invitation and their hospitality. FM acknowledges partial support from the CNRS PEPS-PTI project ``DARKMONO''. 
TH thanks the CERN theory group for its hospitality during his regular visits to CERN.


\end{document}